\documentclass[prb,twocolumn,superscriptaddress,showpacs]{revtex4}
\usepackage{amssymb}
\usepackage{graphicx}
\usepackage[dvips]{color}
\definecolor{MyOrange}{rgb}{1,0.6471,0}
\begin{document}
\title{\bf Sound velocities of hexagonal close-packed H$_2$ and He under pressure}
\author{Yu. A. Freiman}
\email{freiman@ilt.kharkov.ua} \affiliation{B.Verkin Institute for
Low Temperature Physics and Engineering of the National Academy of
Sciences of Ukraine, 47 Lenin avenue, Kharkov, 61103, Ukraine}
\author{Alexei Grechnev}
\affiliation{B.Verkin Institute for Low Temperature Physics and
Engineering of the National Academy of Sciences of Ukraine, 47
Lenin avenue, Kharkov, 61103, Ukraine}
\author{S. M. Tretyak}
\affiliation{B.Verkin Institute for Low Temperature Physics and
Engineering of the National Academy of Sciences of Ukraine, 47
Lenin avenue, Kharkov, 61103, Ukraine}
\author{A. F. Goncharov}
\affiliation{Geophysical Laboratory, Carnegie Institution of
Washington, 5251 Broad Branch Road NW, Washington, DC 20015, USA}
\author{C. S. Zha}
\affiliation{Geophysical Laboratory, Carnegie Institution of
Washington, 5251 Broad Branch Road NW, Washington, DC 20015, USA}
\author{Russell J. Hemley}
\affiliation{Geophysical Laboratory, Carnegie Institution of
Washington, 5251 Broad Branch Road NW, Washington, DC 20015, USA}

\begin{abstract}
{Bulk, shear, and compressional aggregate sound velocities of
hydrogen and helium in the close-packed hexagonal structure are
calculated over a wide pressure range using two complementary
approaches: semi-empirical lattice dynamics based on the many-body
intermolecular potentials and density-functional theory in the
generalized gradient approximation. The sound velocities are used
to calculate pressure dependence of the Debye temperature. The
comparison between experiment and first-principle and
semi-empirical calculations provide  constraints on the density
dependence of intermolecular interactions in the zero-temperature
limit.}

\pacs{67.80.F-,67.80.B-,62.20.dj}
\end{abstract}
\maketitle

Hydrogen and helium are the simplest and most abundant chemical
elements in the Universe. Studies of solid helium and hydrogen at
elevated pressures, which started at the end of the 1920s, are of
great interest for many branches of science. Hydrogen and helium
are major constituents of stars and giant planets and their
physical properties are very important for condensed matter
physics, planetary science and astrophysics. As such the behavior
of these elements under extreme environments of pressure and
temperature is central to modeling the interiors of planetary and
astrophysical bodies.

High-pressure x-ray, Raman and infra-red (IR) studies established
three molecular phases of solid hydrogen
\cite{Silvera80,Kranendonk83,Mao94,Manzhelii97}. These phases are
related to the orientational ordering  of the molecules and
structural transitions. In phase I, which is stable at zero
temperature up to 110 GPa, hydrogen molecules are quantum rotors
arranged in the hexagonal-close-packed (hcp) structure. At the I -
II phase transition the molecules go from the quantum rotor state
to a strongly anharmonic anisotropic librational state. Above 150
GPa solid hydrogen transforms to phase III. The molecular ordering
in this phase can be treated classically. Thus, for this part of
the phase diagram  the most general picture can be formulated in
terms of the concept of quantum versus classical orientational
ordering \cite{Mazin97}. Although the structures of phases II and
III are unknown, x-ray and Raman data suggest that the hydrogen
molecules in both phases lie close to the sites of the hcp lattice
\cite{Akahama10,Freiman12}. At low temperatures the molecules are
stable to at least 360 GPa; but hydrogen transforms to new phases
(e.g., phase IV) with increasing temperature at these pressures
\cite{Eremets11,Howie12,Zha12,Zha13}.

At low pressures and temperatures $^4$He crystallizes into the hcp
structure. High-pressure x-ray diffraction measurements
\cite{Mao88,Loubeyre93,Zha04} have shown that in a wide
temperature (up to 400 K) and pressure (up to 58 GPa) range hcp
$^4$He is stable with the exception of two narrow segments
adjacent to the melting curve (25.9 - 30.4 bar, bcc, and 0.1 -
11.6 GPa, fcc). The highest volume compression reached in the
equation of state (EOS) experiments is $V_0/V=10.4$ at 180 GPa for
solid hydrogen \cite{Akahama10} ($V_0/V=7.6$ for solid D$_2$
\cite{Loubeyre96}), and $V_0/V=8.4$ for solid helium
\cite{Loubeyre93}.

The phonon spectra of hcp hydrogen and helium exhibits  a
Raman-active optical mode  of the $E_{2g}$ symmetry. The frequency
$\nu(P)$ of this mode calculated with various semi-empirical (SE)
potentials is highly sensitive to details of the calculation
scheme, making it a stringent  test for any potential or
theoretical method, e.g., density functional theory (DFT). The
frequency range of this mode in solid hydrogen as a function of
pressure is extremely large: from 36 cm$^{-1}$ at zero pressure
\cite{Silvera72} to ~ 1100 cm$^{-1}$ at ~250 GPa
\cite{Goncharov98,Goncharov01}. For solid helium the situation is
different: neutron measurements performed close to the
solidification pressure \cite{Eckert77} gave about 50 cm$^{-1}$
for the $E_{2g}$-mode frequency; Raman measurements under pressure
of about 1 GPa \cite{Watson85} gave frequency about 74 cm$^{-1}$.
Higher pressure  measurements have not revealed any Raman
activity\cite{Polian86,Goncharov}. According to DFT and
semi-empirical (SE) calculations \cite{Freiman08} at the highest
reached compressions the Raman frequency is about 500 cm$^{-1}$.

Sound velocity is another experimentally measurable quantity which
provides information on the elastic moduli, elastic anisotropy,
equation of state, and other thermodynamic properties. At low
pressures, the lattice dynamics of solid hydrogen and helium is
governed by strongly anharmonic and quantum crystal effects. The
use  of the self-consistent phonon (SCP) approach  made it
possible to reach good agreement between theoretical
\cite{Goldman79a,Goldman79b,Goldman80} and measured sound
velocities and elastic moduli \cite{Wanner73,Thomas78,Nielsen73}
for parahydrogen and orthodeuterium. The situation is similar for
solid helium: there is a good correspondence between early
experimental results  \cite{Frank70,Crepeau71,Greywall71} and SCP
theories \cite{Goldman70,Horner71,Gillis68}. A detailed review of
early literature  for solid helium was given by Trickey {\it et
al.} \cite{Trickey72}.

The problem of supersolid have rekindled interest in experimental
\cite{Day07,Syshchenko09} and theoretical
\cite{Cazorla08,Pessoa10,Cazorla12,Cazorla13} studies of elastic
properties of solid helium in the quantum crystal region. At the
same time, studies of elastic properties of solid hydrogen and
solid helium at elevated pressures are rather scarce. Liebenberg
{\it et al.} \cite{Liebenberg78} measured the sound velocities in
solid hydrogen from 0.4 to 1.9 GPa using the piston-cylinder
technique. Zha {\it et al.} \cite{Zha93} and Duffy {\it et al.}
\cite{Duffy94} studied the elasticity of solid hydrogen in the
pressure range up to 24 GPa by single-crystal Brillouin
scattering.  In solid helium the aggregate quasi-compressional
sound velocity $v_P$ to 20 GPa was found by combing results of two
experiments: Polian and Grimsditch \cite{Polian86} measured the
product of $v_P$ and the refractive index $n$ by using Brillouin
scattering in the backscattering geometry,  and Le Toullec {\it et
al.} made a separate refractive-index measurements
\cite{LeToullec89}. The direct data on the sound velocities in
solid helium up to 32 GPa  were obtained by the single-crystal
Brillouin scattering measurements by Zha {\it et al.}
\cite{Zha04}.

The goal of the present paper is to calculate the pressure
dependence of sound velocities of hcp H$_2$ and He over a wide
range of pressures and compare with existing experimental data. No
explicit effects of electronic excitations or changes in molecular
bonding (e.g,, as a function of temperature) are assumed. As such,
the results provide a baseline for comparison with more elaborate
models (e.g., that go beyond DFT or include thermal effects). The
calculations were carried out using the DFT and semi-empirical
(SE) approaches. The results for hydrogen extend  our previous
results \cite{Freiman13}. The elastic properties of He under
pressure were previously investigated within DFT using an
atomic-based EMTO code \cite{Nabi05}. Unfortunately, the
small-scale figures shown in Ref. 48 render a quantitative
comparison, in particular, with the SE results difficult. The
comparison of the SE and DFT-GGA results shows that these two
approaches complement each other \cite{Freiman12}: at lower
pressures SE gives more accurate results but with increasing
pressure DFT-GGA becomes preferable.

The hydrodynamic or bulk sound velocity $v_B$ was found from the
EOS:
\begin{equation}
\label{v_B} v_B = [\partial P/\partial \rho]^{1/2}= \left[-\,
\frac{V^2}{\mu}\frac{\partial P}{\partial V}\right]^{1/2},
\end{equation}
where $P$ is the pressure,  $\rho$ is the density, $\mu$ is the
molar mass, and $V$ is the molar volume.

The shear velocity $v_S$ was obtained from the the shear elastic
constant $C_{44}$, which was in turn calculated from the Raman
frequency $\nu(E_{2g})$ using the relation \cite{Olijnyk00}:
\begin{equation}
\label{v_S} \nu(E_{2g})=
\left(4\sqrt{3}a^2C_{44}/(mc)\right)^{1/2},
\end{equation}
where $a,\,c$ are the lattice parameters and $m$ is the molecular
mass. This relation was used in Ref. \onlinecite{Olijnyk00} for
solid He under pressure.

In the framework of the Voigt-Reuss-Hill averaging scheme
\cite{Voigt1910,Reuss1929,Hill1952,Korpiun,Steinle} the relation
between the isotropically averaged aggregate compressional, $v_P$,
bulk, $v_B$, and shear, $v_S$, sound velocities holds
\cite{Anderson63}
\begin{equation}
\label{v_P} v_P^2 = v_B^2+\frac{4}{3}v_S^2.
\end{equation}
Using this equation we calculated the aggregate compressional
sound velocity $v_P$ which together with $v_B$ and $v_S$ made it
possible to calculate the Debye temperature $\Theta_D$:
\begin{equation}
\label{T_D} \Theta_D=\frac{\hbar}{k_B}
\left[\frac{V}{18\pi^2N_A}\left(\frac{1}{v_P^3}+\frac{2}{v_S^3}\right)\right]^{-1/3},
\end{equation}
where $k_B$ is the Boltzmann constant and $N_A$ is the Avogadro
number.

We start with the DFT calculations. The sound velocities $v_B$,
$v_S$ in H$_2$ and He were calculated within DFT using the
generalized gradient approximation (GGA) \cite{PBE}. All
calculations were done using the full-potential linear muffin-tin
orbital (FP-LMTO) code RSPt \cite{wills-book} for zero
temperature. The $Pca{2\rm_1}$ structure was used for hydrogen.
The zero-point vibrations (ZPV) of the nuclei were ignored in the
initial calculations.

\begin{figure}
\includegraphics[scale=0.3]{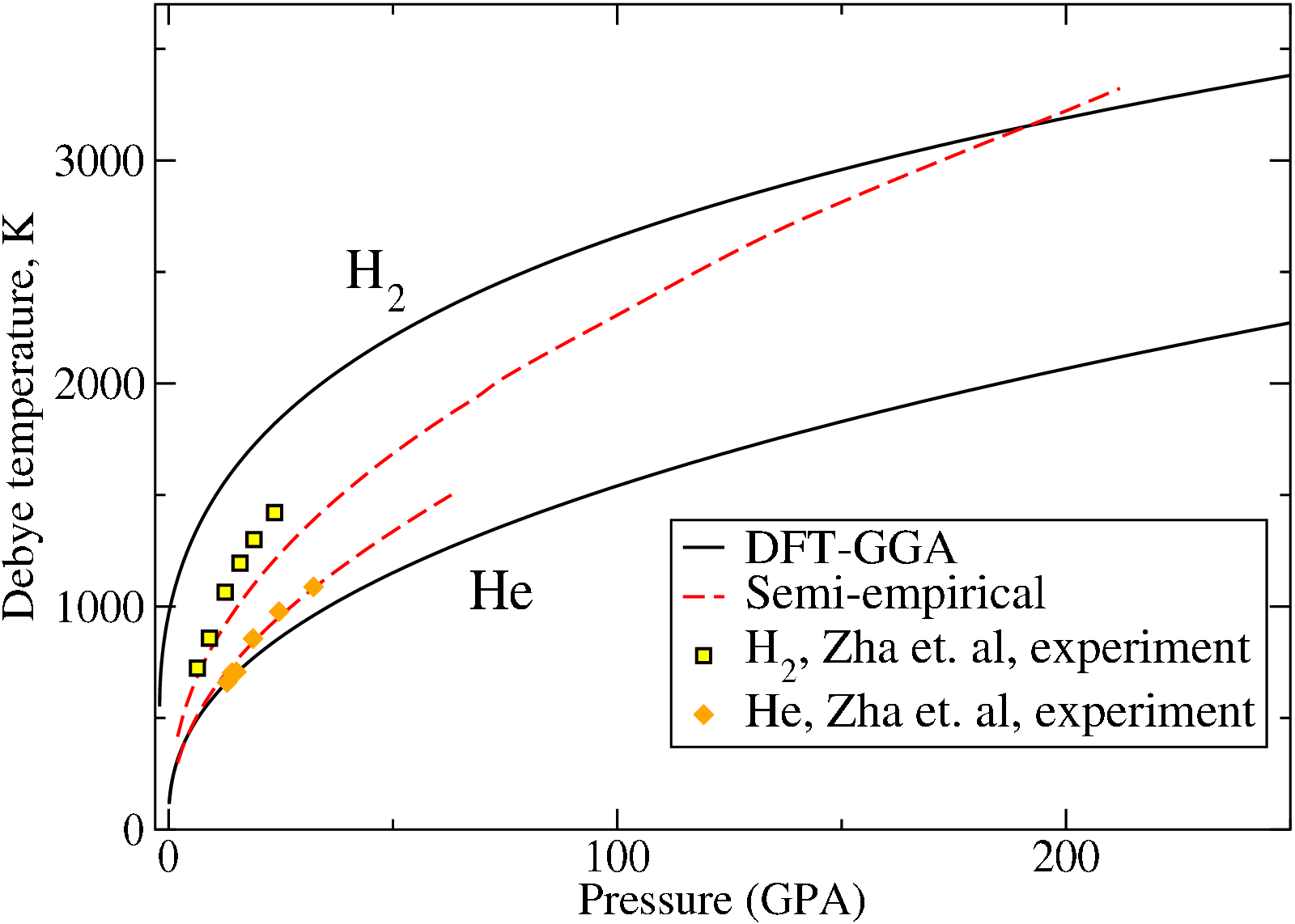}
\caption{\label{f_T_D} (Color online) Debye temperature in solid
H$_2$ and He as a function of pressure. Theoretical many-body SE
and DFT-GGA results (this work), and experimental data for
H$_2$\cite{Zha93} and He \cite{Zha04} are presented.}
\end{figure}

The bulk sound velocities were calculated from the parametrized
DFT EOS $P(V)$ by numerical differentiation (Eq. (\ref{v_B})). The
shear sound velocities were found from the DFT Raman frequencies
(Eq. (\ref{v_S})), which were in turn obtained from supercell
total energy calculations and parametrized to allow for a smooth
numerical differentiation of the Debye temperature. 693 and 1331
$k$-points in the Brillouin zone were used for H$_2$ and He,
respectively, and the convergence of the results with the number
of $k$-points was checked. GGA EOS was used to obtain
$P$-dependent sound velocities from the $V$-dependent ones. As a
result, from Eqs. (\ref{v_B}) - (\ref{v_P}) we obtained the
zero-order aggregate sound velocities and from Eq. (\ref{T_D}) we
obtained the Debye temperature $\Theta_D $ (Fig. \ref{f_T_D}).

The ZPV were  taken into account in our DFT-GGA approach within
the framework of the Debye model. The ZPV correction to the EOS is
\begin{equation}
\Delta P(V) = - \frac 98 N_A k_B \frac {d \Theta_D}{dV},
\end{equation}

This formula was used to calculate the ZPV-corrected $P(V)$ and
$v_B(V)$ from the original (non-ZPV-corrected) $v_B(V)$, $v_S(V)$.
The shear velocity $v_S(V)$ is not affected by ZPV in our
approximation. Finally, the ZPV-corrected EOS is used to calculate
the ZPV-corrected $P$-dependent velocities $v_B(P)$ and $v_S(P)$.

\begin{figure}
\includegraphics[scale=0.29]{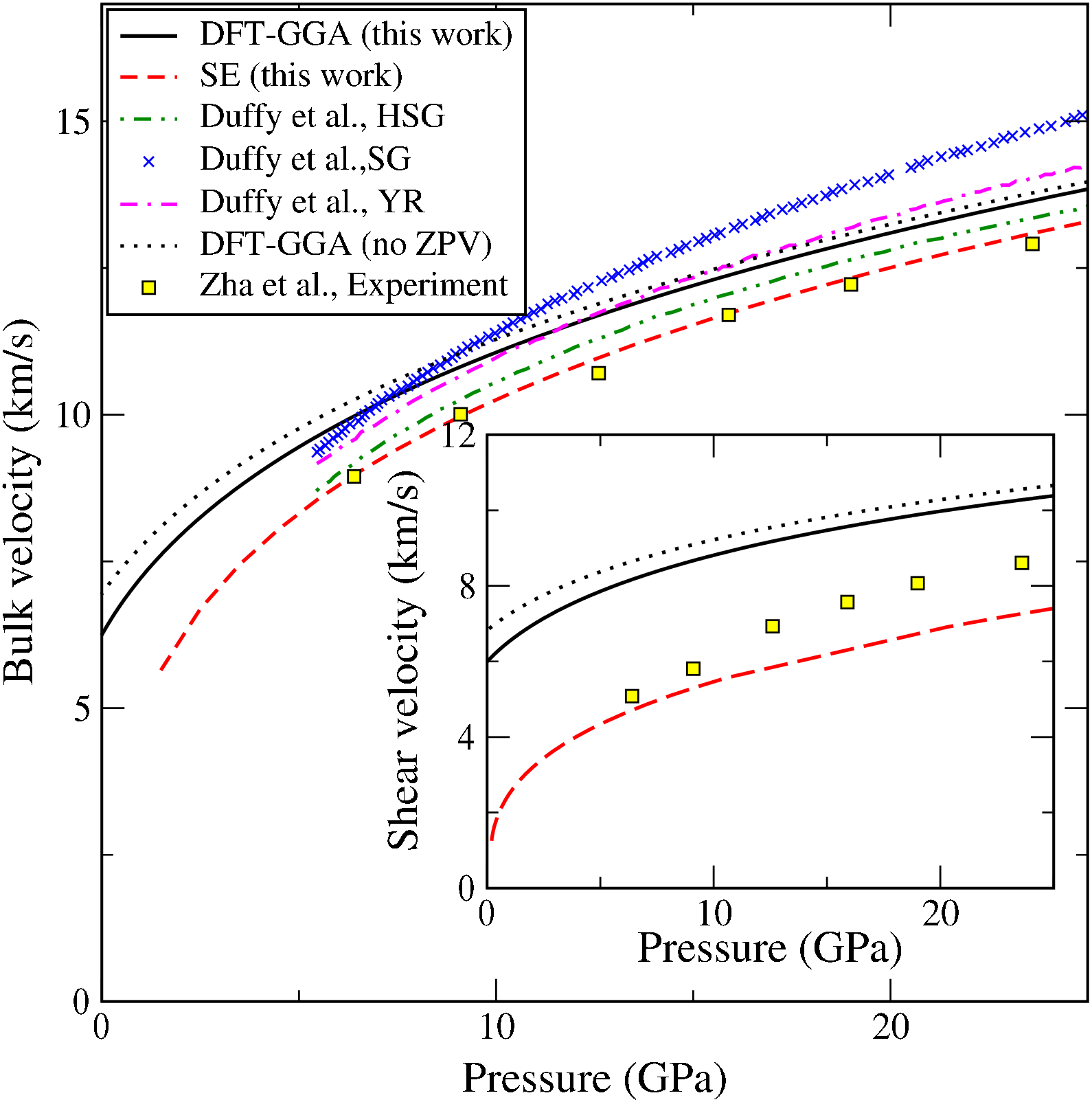}
\caption{\label{f_h2} (Color online) Bulk sound velocity in solid
H$_2$ as a function of pressure.  Theoretical results obtained
using many-body SE and DFT-GGA including  and disregarding ZPV
(this work), SE with pair potentials SG \cite{Duffy94}, HSG
\cite{Duffy94} and YR \cite{Duffy94}, and the experimental data
\cite{Zha93} are presented. The inset shows the SE and DFT-GGA
shear sound velocities (this work) in comparison with experiment
\cite{Zha93}.}
\end{figure}

\begin{figure}
\includegraphics[scale=0.285]{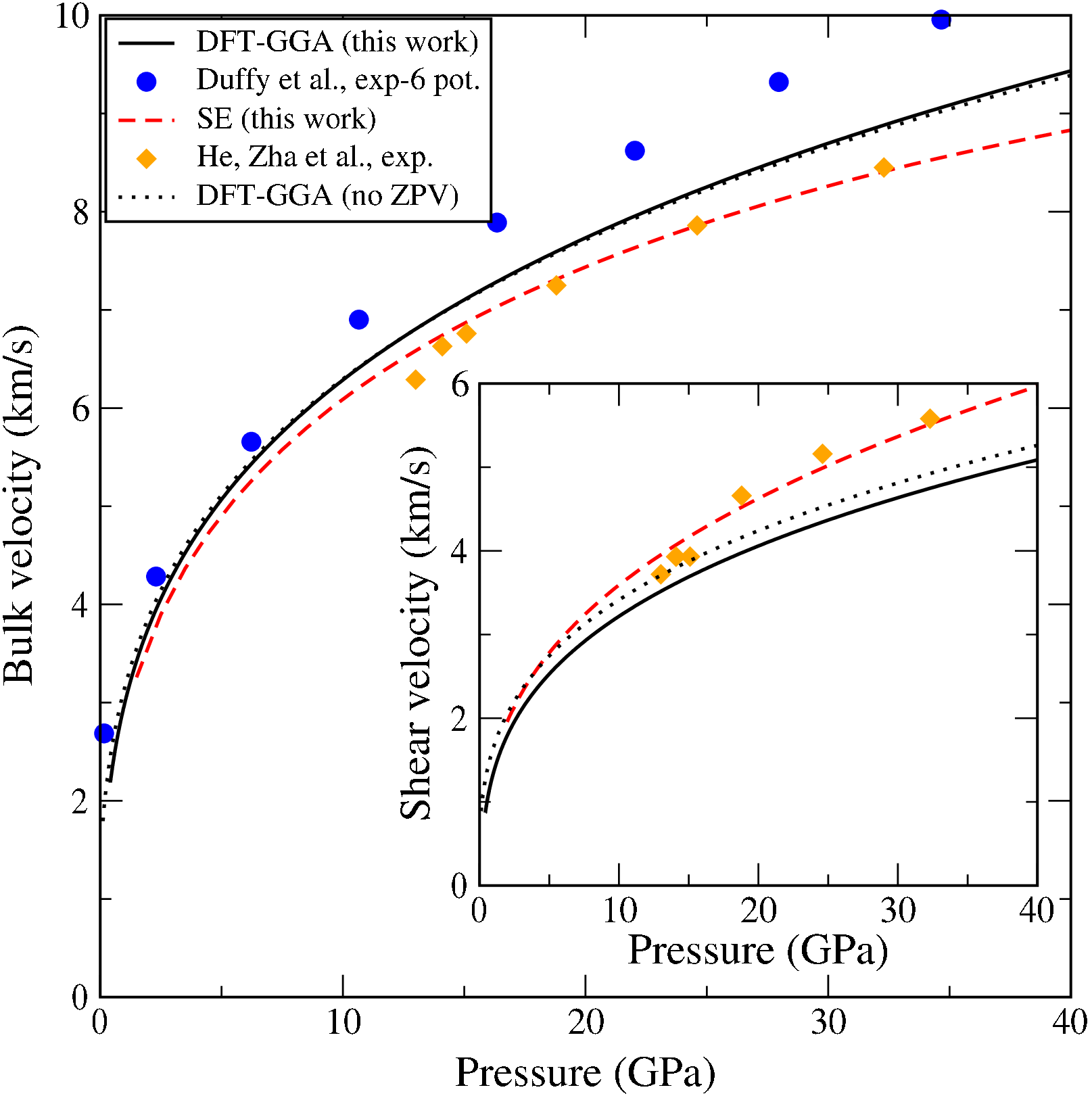}
\caption{\label{f_he} (Color online) Bulk sound velocity in solid
He as a function of pressure. Theoretical results obtained using
many-body SE and DFT-GGA including and disregarding ZPV (this
work), SE with exp - 6 pair potential \cite{Duffy94}, and the
experimental data \cite{Zha04} are presented. The inset shows the
SE and DFT-GGA shear sound velocities (this work) in comparison
with experimental results \cite{Zha04}.}
\end{figure}

We now turn to the SE calculations. A variety of pair potentials
have been tested in EOS and Raman studies of solid hydrogens
\cite{Silvera78,Ross83,Hemley90,Duffy94}. One of the first was the
low-pressure Silvera - Goldman potential (SG) \cite{Silvera78}.
The first DAC  experiments found it to be too stiff, i.e. the
repulsion increases too rapidly with pressure. Hemley {\it et al.}
\cite{Hemley90,Duffy94} modified the SG potential \cite{Silvera78}
with a short-range correcting term. This Hemley-Silvera-Goldman
effective potential (HSG) as well as other pair potentials, e.g.
the Young-Ross (YR) potential \cite{Ross83}, were shown to fit the
experimental EOS well up to 40 GPa, but they are still too stiff
at yet higher pressures \cite{Loubeyre96}. A similar situation
takes place for helium \cite{Loubeyre93,Freiman07}.  The reason is
the neglect of the three body and higher-order terms in the
intermolecular potential
\cite{Loubeyre87,Freiman07,Freiman08,Freiman09,Grechnev10,Freiman11,Freiman12}.
\begin{figure}
\includegraphics[scale=0.28]{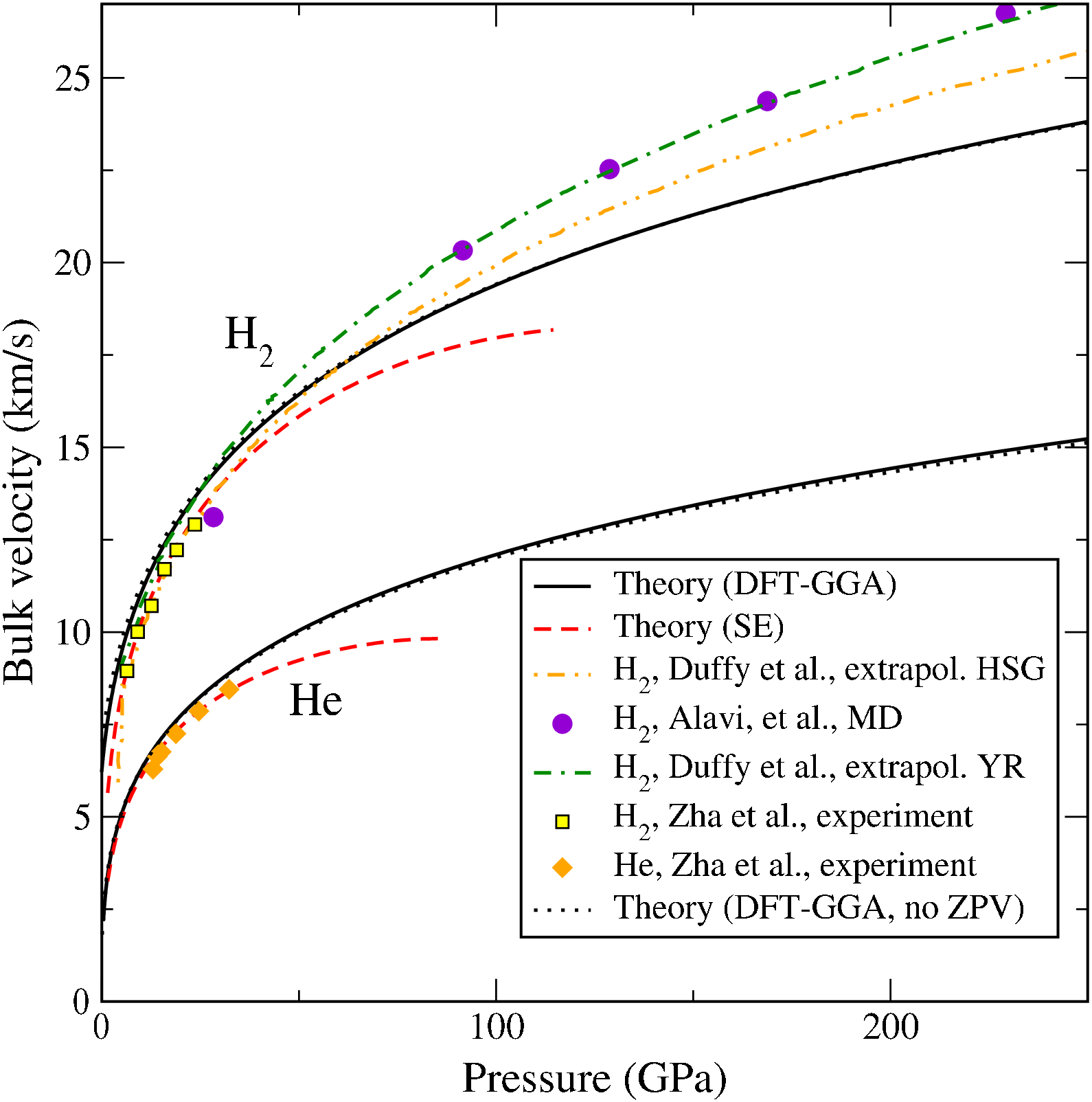}
\caption{\label{f_Bulk} (Color online) Bulk sound velocity in
solid H$_2$ and He as a function of pressure for the extended
pressure range. Theoretical  many-body SE and DFT-GGA results
including and disregarding ZPV (this work), first-principle MD
results \cite{Alavi95}, SE with HSG and exp - 6 pair potentials
\cite{Duffy94}, and experimental data \cite{Zha04,Zha93} are
presented.}
\end{figure}
\begin{figure}
\includegraphics[scale=0.28]{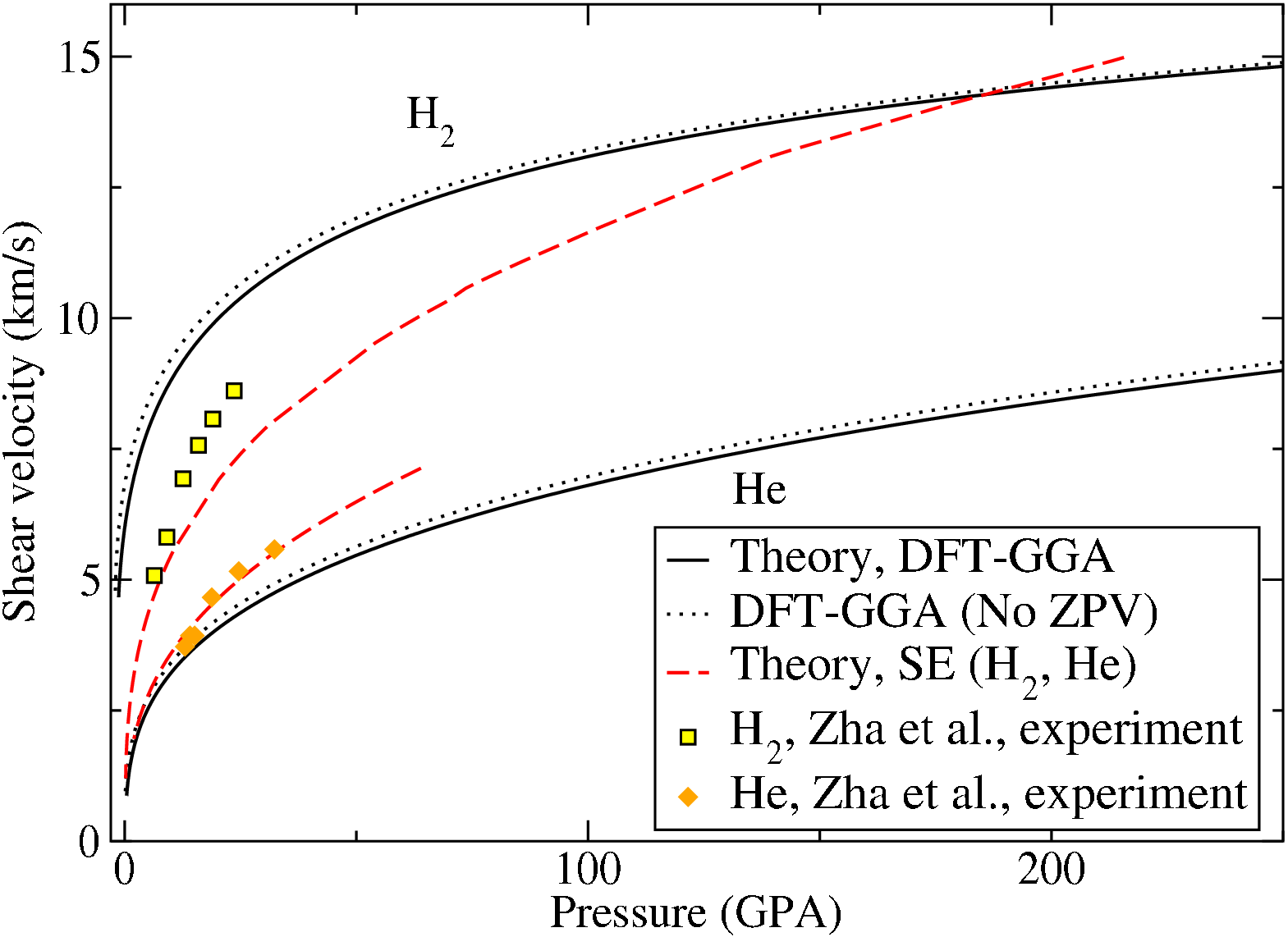}
\caption{\label{f_Shear} (Color online) Shear sound velocities of
hcp hydrogen and helium for the extended pressure range.
Theoretical many-body SE and DFT-GGA results including and
disregarding zero-point vibrations (ZPV) (this work), and
experimental data \cite{Zha04,Zha93} are presented.}
\end{figure}
First-principles methods and SE approaches which take into account
such terms work well up to the highest pressures reached in EOS
and Raman measurements \cite{Freiman12}.

The many-body hydrogen intermolecular potential  used here is a
sum of the pair SG potential \cite{Silvera78} (discarding the
$R^{-9}$ term) and two three-body terms: the long-range
Axilrod-Teller dispersive interaction and the short-range
three-body exchange interaction in the Slater-Kirkwood form
\cite{Loubeyre87,Freiman08}. The explicit form and parameters of
the potential used in this work for solid hydrogen are given in
Ref. \onlinecite{Freiman11}. The interatomic potential for solid
helium has a similar form \cite{Loubeyre87,Freiman07,Freiman08}.
For the two-body interaction, we used the HFDHE2 Aziz {\it et al.}
potential \cite{Aziz79}.  We also tested the HFD-B3-FCI1 Aziz
potential \cite{Aziz95} and found that the results for these two
pair potentials practically coincide.  The explicit form of the
potential used for solid helium are given in Ref.
\onlinecite{Freiman07}. In our calculations we restrict ourselves
to $T = 0$ K, with the zero-point energy treated in the Einstein
approximation.

The calculated bulk and shear sound velocities for H$_2$  and He
are shown in Figs. (\ref{f_h2}) and (\ref{f_he}) for solid H$_2$
and He, respectively. The many-body SE bulk velocities are in
excellent agreement with the data from Brillouin scattering
measurements \cite{Zha93,Duffy94,Zha04}. Since the empirical SG
\cite{Silvera72} and HSG \cite{Hemley90} potentials for H$_2$ tend
to overestimate the repulsive part of the intermolecular
interaction, they also underestimate the compressibility  and
overestimate the sound velocity. Thus, the stiffer the potential
is, the greater the error. The same conclusion follows from the
comparison of the bulk sound velocity in solid He calculated from
the many-body potential and far more rigid exp-6 potential (Fig.
\ref{f_he}). As for DFT-GGA results, in the pressure range shown
in Figs. \ref{f_h2} and \ref{f_he} their accuracy are markedly
below that of the SE results which reflects the fact that DFT-GGA
does not treat the van der Waals interaction properly.

In order to make a proper comparison of the results of the both
approaches we calculated the sound velocities over a broad range
of pressures in the ideal hcp structure. Figures (\ref{f_Bulk})
and (\ref{f_Shear}) present the bulk and shear sound velocities,
respectively, in solid H$_2$ and He calculated with our SE and
DFT-GGA methods for the extended pressure range up to 250 GPa. The
first-principles molecular dynamics (MD) results \cite{Alavi95}
for H$_2$ by Alavi {\it et al.}, and the SE results by Ross {\it
et al.} \cite{Ross83} (HSG and YR potentials) extrapolated by
Duffy {\it et al.} \cite{Duffy94}, are also presented for
comparison. Although the MD results by  Alavi {\it et al.} agree
well with the extrapolated YR results, they are still
significantly higher than our SE and DFT-GGA results.

A rather narrow range of pressures where experimental results
\cite{Zha93,Zha04} exist did not permit unambiguous bounds on the
of applicability of the DFT-GGA  and SE approaches used in the
sound velocity calculations. Judging from Figs. (4) and (5), upper
bounds for the SE approach are approximately  75 GPa and
approximately 50 GPa for H$_2$ and He, respectively, and the lower
bound for the DFT-GGA is around 100 GPa both for H$_2$ and He. So
the upper bound for the SE and the lower bound for the DFT-GGA do
not overlap and there is an intermediate  pressure range where
both methods are ineffective.  It should be noted that the ranges
of applicability of the SE approach are different for the EOS
\cite{Freiman12}, Raman \cite{Freiman12}, and sound velocity
calculations.

In conclusion, the bulk and shear sound velocities in solid H$_2$
and He calculated with the SE many-body intermolecular potential
are in good agreement with experiment. At low pressures the
accuracy of the SE approach is much higher than that of the
DFT-GGA, but for pressures over 100 GPa the latter approach is
preferable. The results can serve as a baseline for planetary and
astrophysical models and provide  a basis for extrapolation to
more extreme conditions.

The work was supported by the  National Nuclear Security
Administration, Centre for Development of Advanced Computing, and
NSF DMR 1106132. A.F.G. acknowledges support from the NSF, Army
Research Office, NASA Astrobiology Institution, and Energy
Frontier Research in Extreme Environments.

\end{document}